\documentclass[10pt,oneside,onecolumn,preprint,number,sort,compress]{elsarticle}
\pdfoutput=1
\usepackage{amsmath}
\usepackage{amssymb}
\newcommand{\PRE}[1]{}       % Use if journal style

\usepackage{enumerate}
\usepackage{url}
\usepackage{color}
\usepackage{xcolor}
\usepackage{slashed}
\usepackage{subfigure}
\usepackage{rotating}
\usepackage{epsfig}
\usepackage{bm}
\usepackage{epstopdf}
\usepackage{graphicx}
\usepackage[colorlinks=true,citecolor=blue,urlcolor=blue]{hyperref}
\renewcommand{\eqref}[1]{Eq.~(\ref{#1})}

\newcommand{\nnbar}{n-{\bar n}} 

\newcommand{\DeltaBm}{|\Delta {\rm B}|}
\newcommand{\DeltaLm}{|\Delta {\rm L}|}

\begin{document}

\begin{frontmatter}
\title{Processes that break baryon number by two units 
and the Majorana nature of the neutrino}

\author[val]{Susan Gardner}\ead{gardner@pa.uky.edu}
\author[val]{Xinshuai Yan}\ead{xinshuai.yan@uky.edu}
\address[val]{Department of Physics and Astronomy, University of 
Kentucky, Lexington, Kentucky 40506-0055 USA}

%%%%%%%%%%%%%%%%%%%%%%%%%%%%%%%%%%%%%%%%%%%%%%%%%%%%%%%%%%%%%%%%%%%%%%%%%%%%%%%
\begin{abstract}
We employ the simplest possible models of scalar-fermion interactions that
are consistent with the gauge symmetries of the Standard Model and permit no proton 
decay to analyze the connections possible among processes that break baryon number
by two units. In this context we show how the 
observation of $n$-${\bar n}$ oscillations and 
of a pattern of particular nucleon-antinucleon conversion processes --- all accessible
through e-d scattering --- namely, selecting from 
$e^- p \to e^+ {\bar p}$, 
$e^- p \to {\bar n} {\bar \nu}$,  
$e^- n \to {\bar p} {\bar \nu}$, and 
$e^- n \to e^- {\bar n} $ 
would reveal that the decay $\pi^- \pi^- \to e^- e^-$ must occur also. 
This latter process is the leading contribution 
to neutrinoless double beta decay in nuclei 
mediated by new short-distance physics, in contrast to that mediated 
by light Majorana neutrino exchange. 
The inferred existence of $\pi^- \pi^- \to e^- e^-$ 
would also reveal the Majorana nature of the neutrino, 
though the absence of this inference would not preclude it. 
\end{abstract}
%\pacs{11.30.Fs, 14.70.Pw, 12.60.Cn, 13.60.-r, 23.40.Bw}

\end{frontmatter}
%14.70.Pw Other gauge bosons
%12.60.Cn Extensions of electroweak gauge sector
%13.60.-r Photon and charged-lepton interactions with hadrons (for neutrino interactions, see 13.15.+g)
%23.40.Bw Weak-interaction and lepton (including neutrino) aspects (see also 14.60.Pq Neutrino mass and mixing) (beta decay)
%11.30.Fs Global symmetries (e.g., baryon number, lepton number)

%\section{Introduction}
%\label{sec:intro}
\section{Introduction}
The quantity baryon number (B) - lepton number (L), B-L, 
is exactly conserved in the Standard Model (SM), so that 
the observation of B-L violation would reveal the existence of new physics. 
In this letter we consider the possibility of the discovery of B-L violation within the
realm of the strong interactions and the quark sector --- 
and its broader implications. We focus particularly on processes that 
break baryon number by two units because proton decay, or, more generally, processes
with $\DeltaBm =1$, are not only unobserved but also have exceptionally 
strong empirical limits on their non-existence~\cite{Patrignani:2016xqp}. 
Moreover, as long known, 
the new-physics origins of $\DeltaBm =1$ and $\DeltaBm =2$ processes can be completely
distinct~\cite{Mohapatra:1980qe,Babu:2001qr,Arnold:2012sd,Dev:2015uca}. 

The prospect of B-L violation is often discussed in the context of the fundamental nature of
the neutrino; its violation would both make the $\DeltaLm=2$ process of neutrinoless 
double beta ($0\nu \beta\beta$) decay possible and give the 
neutrino a Majorana mass~\cite{Weinberg:1979sa,Chikashige:1980ui,Schechter:1981cv}, 
revealing that the neutrino can be regarded as its own antiparticle~\cite{Schechter:1981bd}. 
General parametrizations of the decay rate are associated with 
the long-range exchange of a light Majorana 
neutrino~\cite{Pas:1999fc,Deppisch:2012nb,Helo:2016vsi}, 
or through a short-range process mediated by new B-L violating dynamics at roughly the TeV 
scale~\cite{Pas:2000vn}. 
The nuclear matrix elements, which are needed
to interpret $0\nu \beta\beta$ experiments, 
differ considerably 
in the two cases~\cite{Haxton:1985am,Pas:1999fc,Pas:2000vn,Vergados:2012xy,deGouvea:2013zba}. 
Systematic analyses of the possible 
operators 
of $0\nu \beta\beta$ decay~\cite{Babu:2001ex,deGouvea:2007qla,Bonnet:2012kh,Graesser:2016bpz} 
and of 
the associated decay topologies~\cite{Bonnet:2012kh}, and of the decay rate within 
 chiral effective theory~\cite{Cirigliano:2017ymo,Cirigliano:2017djv,Cirigliano:2018yza}
exist. The short-range mechanism is captured by $\pi^- \pi^- \to e^- e^-$~\cite{Vergados:1981bm}
at leading order in hadron chiral effective theory~\cite{Prezeau:2003xn}, 
and the size of the associated hadronic matrix element has recently been computed
in lattice QCD~\cite{Nicholson:2018mwc}. 
We believe that insight on the mechanisms of $0\nu \beta\beta$ decay can be gleaned
through the study of B-L violation in the quark sector, as it is the short-distance mechanism
that can connect B-L violation with quarks to that with leptons. 

The empirical study of $\DeltaBm=2$ processes has traditionally been associated with 
the search for $\nnbar$ oscillations 
with free or bound nucleons~\cite{kuzmin1970zhetf,Glashow1979,Mohapatra:1980qe,mohapatra1980pheno}  and dinucleon decay
in nuclei~\cite{Kabir:1983qx, Basecq:1983hi, Arnold:2012sd,Dev:2015uca,Berger:1991fa, Bernabei:2000xp, Litos:2014fxa, Gustafson:2015qyo}. 
Recently we have proposed the study of 
$\nnbar$ conversion~\cite{Gardner:2016wov, Gardner:2017szu}, 
which, in contrast 
to $\nnbar$ oscillation, would not be spontaneous but 
mediated by an external source. In this letter we discuss the connections between
these possibilities in the context of simple models of B and B-L violation. 
Motivated by ``minimal'' models for connectors to 
new hidden sectors~\cite{Bjorken:2009mm,Batell:2009di}, 
we introduce
new scalar gauge bosons whose interactions are of mass dimension 3 and 4, so that 
the new interactions can be added to the SM in a theoretically consistent way. 
Scalar-fermion interactions in such models that respect the gauge 
symmetries of the SM have been studied in some 
detail~\cite{Davies:1990sc,Arnold:2012sd,Arnold:2013cva}. 
In the current case our interest is in the models that permit 
$\DeltaBm =2$ transitions without
proton decay, and indeed in those that do not permit 
$\DeltaBm =1$ transitions~\cite{Arnold:2012sd,Dev:2015uca}. 
% at tree level. 
%Models along these lines exist
%, and, 
Interestingly, we have discovered 
that a variant of the models of Arnold, Fornal, and Wise~\cite{Arnold:2012sd}
 can be used to generate a $\DeltaLm =2$ transition, 
particularly, that of $\pi^-\pi^- \to e^- e^-$, whose
existence drives the appearance of $0\nu \beta$ $\beta$ decay if mediated
by new short-distance physics~\cite{Prezeau:2003xn}. 
Thus in what follows we consider not only 
how particular $\nnbar$ oscillation and conversion processes can appear in these models, 
but we also show how such models 
can give rise to $0\nu \beta \beta$ decay in nuclei
 --- and we consider the interconnections between them.
Particularly, we discuss how possible patterns of discovery of 
$\DeltaBm=2$ processes can reveal
whether the short-distance dynamics 
that could give rise to $\pi^-\pi^- \to e^- e^-$ can 
be shown to exist. In contrast, Babu and Mohapatra have shown that in the case
of the SO(10) grand unified theory --- 
and independently from the expected existence of the SM sphaleron --- that if 
$\nnbar$ oscillations and a $\DeltaBm =1$ process 
were observed to occur that 
one could also conclude the existence of a Majorana neutrino~\cite{Babu:2014tra}. 
Here we show that such a connection can be demonstrated without requiring the
observation of proton decay, or indeed of any $\DeltaBm =1$ process. 
%We emphasize that in this case, as in Ref.~\cite{Babu:2014tra}, the existence of 
%such an inference does not imply that the mass mechanism 
%ought saturate the experimental rate for $0\nu \beta\beta$ decay. 
We emphasize that in this case, as in Ref.~\cite{Babu:2014tra}, the existence of 
such an inference does not imply that the short-distance mechanism 
ought saturate the experimental rate for $0\nu \beta\beta$ decay. 
Our approach, however, is different from that of Ref.~\cite{Babu:2014tra}, 
as it relies on the use of minimal scalar models. 

\section{Minimal scalar models with baryon number violation but no proton decay}
%\label{scalar}
The minimal scalar models that give rise to $\DeltaBm=2$ and not $\DeltaBm=1$ processes 
while respecting SM gauge symmetries 
contain either three or four scalar interactions. 
Following Refs.~{\cite{Davies:1990sc,Bowes:1996xy,Arnold:2012sd,Arnold:2013cva}} 
we consider all the interactions permitted by Lorentz and 
SU(3)${}_{\rm c}\times$SU(2)${}_{\rm L}\times $U(1)${}_{\rm Y}$ gauge symmetry. 
Models for processes with both $\DeltaBm=1,2$ have been 
constructed~\cite{Bowes:1996xy,Barr:2012xb,Arnold:2012sd,Arnold:2013cva}, 
though in this paper we follow Ref.~\cite{Arnold:2012sd}.  
The particular scalars that allow B or L violation
to appear  
but do not admit $\DeltaBm=1$ processes at tree level are enumerated in 
Table~\ref{table:scalars}. We have also noted the schematic interactions of the scalars $X_i$
to right-handed leptons and quarks of generation $a$ as $e^a$ and $u^a$, $d^a$
and to left-handed leptons and quarks as $L^a$ and $Q^a$, respectively. 
The symmetries of the scalar 
representations under color SU(3) and/or weak isospin SU(2) can fix the 
symmetry of the associated coupling constant under $a,b$ interchange, and 
we have noted that as well in Table~\ref{table:scalars} --- the relation 
$g_i^{ab} = \pm g_i^{ba}$ indicates $S (+)$ or $A (-)$, respectively, and ``--'' 
denotes no interchange symmetry. 
We note that $X_9$ cannot generate a B and/or L violating
interaction of mass dimension four or less, so that 
 we do not consider it further, 
and that interactions denoted by ``A'' cannot involve only first-generation fermions. 

\begin{table}[tb]
 \caption{Scalar particle representations in the 
SU(3)${}_{\rm c}\times$SU(2)${}_{\rm L}\times $U(1)${}_{\rm Y}$ SM that carry nonzero 
B and/or L but permit no proton decay at tree level, after Ref.~\cite{Arnold:2012sd}. 
We indicate the possible interactions between the scalar $X$ and SM fermions schematically. 
Note that the indices $a,b$ run over three generations, that the symmetry of the associated
coupling $g_i^{ab}$ under $a \leftrightarrow b$ exchange is noted in brackets, and finally that 
our convention for $Y$ is $Q_{\rm em}=T_3 +Y$. Please refer to the text for further discussion.} 
% \vspace*{-.1in}
 \label{table:scalars}
\begin{tabular}[t]{llcclc}
\hline
\hline
Scalar & SM Representation &  \ B \ & \ L \ & \ Operator(s) & [$g_i^{ab}$?] \\
\hline
$X_1$ & $(1, 1, 2)$ & 0 & -2 & $Xe^ae^b\;$ & [S] \\ 
$X_2$ & $(1, 1, 1)$ & 0 & -2 & $XL^aL^b\;$ & [A] \\
$X_3$ & $(1, 3, 1)$ & 0 & -2 & $XL^aL^b\;$ & [S] \\
$X_4$ & $({\bar 6}, 3, -1/3)$ & -2/3 & 0 & $XQ^aQ^b\;$ & [S] \\
$X_5$ &  $({\bar 6}, 1, -1/3)$ & -2/3 & 0 & $XQ^aQ^b, Xu^ad^b\;$ & [A,--]  \\
$X_6$ & $(3, 1, 2/3)$ & -2/3 & 0 & $Xd^ad^b\;$ & [A] \\
$X_7$ & $({\bar 6}, 1, 2/3)$ & -2/3 & 0 & $Xd^ad^b\;$ & [S] \\
$X_8$ & $({\bar 6}, 1, -4/3)$ & -2/3 & 0 & $Xu^a u^b\;$ & [S] \\
$X_9$ & $(3, 2, 7/6)$ & 1/3 & -1 & $X{\bar Q}^ae^b, XL^a{\bar u}^b\;$ & [--,--] \\
%\toprule
\hline
\hline
 \end{tabular}
\vspace*{-.2in}
\end{table}

In what follows we extend the models of Ref.~\cite{Arnold:2012sd} to include
the possibility of $\DeltaLm=2$ processes as well. That earlier work 
focused on the possibility of 
$\DeltaBm=2$ processes without proton decay as mediated by interactions of
the form $X_a^2 X_b$ or $X_a^3 X_b$, where $X_a$ and $X_b$ are simply two distinct scalars
that yield the SM gauge invariant interactions indicated, 
because 
it turns out not to be possible to add just one scalar and achieve that end. 
Here we enumerate all the possible B and/or L violating interactions that appear
in mass dimension of four or less without regard to the number of different
scalars that can appear. With three different scalars we can produce 
$\DeltaLm=2$ processes that also couple to quarks, 
and we study the connections between $\DeltaBm=2$ and
$\DeltaLm=2$ processes explicitly. 

We begin by fleshing out the precise interactions indicated in 
Table~\ref{table:scalars}. Specifically, the possible scalar-fermion interactions
mediated by each $X_i$ are 
\begin{alignat}{5}
&-g_1^{ab} X_1 (e^a e^b) \,, \, 
&&-g_2^{ab} X_2 (L^a \varepsilon L^b) \,, 
&& -g_3^{ab} X_3^A (L^a \xi^A L^b) \,, \, \nonumber \\
&-g_4^{ab} X_4^{\alpha \beta A} (Q_\alpha^a \xi^A Q_\beta^b) \,, 
&&-g_5^{ab} X_5^{\alpha \beta} (Q_\alpha^a \varepsilon Q_\beta^b) \,,\, 
&&-g_5^{\prime ab} X_5^{\alpha \beta} (u_\alpha^a d_\beta^b) \,,\nonumber \\
&-g_6^{ab} X_{6 \alpha} (d_\beta^a d_\gamma^b) \varepsilon^{\alpha\beta\gamma} \,,\, 
&&-g_7^{ab} X_7^{\alpha\beta} (d_\alpha^a d_\beta^b) \,,\, 
&& -g_8^{ab} X_8^{\alpha\beta} (u_\alpha^a u_\beta^b) \,,
\end{alignat}
where $\varepsilon=i\tau^2$ is a totally antisymmetric tensor, 
$\xi^A\equiv ((1+ \tau^3)/2, \tau^1 / \sqrt{2}, (1 - \tau^3)/2)$, and 
 $\tau^A$ are Pauli matrices with $A\in 1,2,3$. 
We note $\varepsilon \tau^A$ was used
in place of $\xi^A$ in Ref.~\cite{Arnold:2012sd}, but that choice couples a single component of 
the scalar weak triplet 
to fermion states of differing total 
electric charge, 
incurring couplings that break electric charge conservation. 
The Greek indices are color labels, and we employ
the SU(3) notation of Ref.~\cite{Cheng:1985bj} for fundamental and complex
conjugate representations. 
We adopt 2-spinors such that the fermion 
products in parentheses are Lorentz invariant, and 
we map to 4-spinors via $(u_{L,R \alpha} d_{L,R \beta}) \rightarrow 
(u_\alpha^T C P_{L,R} d_\beta)$ where $C=i\gamma^0 \gamma^2$ and 
$P_{L,R} = (1\mp \gamma_5)/2$ in Weyl representation~\cite{Dreiner:2008tw}. 

\begin{table}[tb]
 \caption{Minimal interactions 
that break B and/or L from scalars $X_i$ that do not permit $\DeltaBm=1$ 
interactions at tree level, indicated schematically, with the Hermitian conjugate implied. 
Interactions labelled M1-M9 appear in models 1-9 of 
 Ref.~\cite{Arnold:2012sd}. Interactions A-G possess $\DeltaLm=2$, $\DeltaBm=0$.
M19, M20, and M21 follow from M8, M17, and M18 
 under $X_7 \rightarrow X_6$, respectively, but they do not
involve 
first-generation fermions only. }
% \vspace*{-.1in}
 \label{table:Xprods}
\begin{tabular}[t]{lccclc}
% \toprule 
\hline
\hline
Model &   &  Model  &  & Model &  \\ 
\hline
M1 & $X_5 X_5 X_7$ &  A  & $X_1 X_8 X_7^\dagger$ & M10 & $X_7 X_8 X_8 X_1$ \\ 
M2 & $X_4 X_4 X_7$ &  B  & $X_3 X_4 X_7^\dagger$ & M11 & $X_5 X_5 X_4 X_3$ \\ 
M3 & $X_7 X_7 X_8$ &  C  & $X_3 X_8 X_4^\dagger$ & M12 & $X_5 X_5 X_8 X_1$ \\ 
M4 & $X_6 X_6 X_8$ &  D  & $X_5 X_2 X_7^\dagger$ & M13 & $X_4 X_4 X_5 X_2$ \\ 
M5 & $X_5 X_5 X_5 X_2$ &  E  & $X_8 X_2 X_5^\dagger$ & M14 & $X_4 X_4 X_5 X_3$ \\ 
M6 & $X_4 X_4 X_4 X_2$ &  F  & $X_2 X_2 X_1^\dagger$ & M15 & $X_4 X_4 X_8 X_1$ \\ 
M7 & $X_4 X_4 X_4 X_3$ &  G  & $X_3 X_3 X_1^\dagger$ & M16 & $X_4 X_7 X_8 X_3$ \\ 
M8 & $X_7 X_7 X_7 X_1^\dagger$ &    &  & M17 & $X_5 X_7 X_7 X_2^\dagger$ \\ 
M9 & $X_6 X_6 X_6 X_1^\dagger$ &    &  & M18 & $X_4 X_7 X_7 X_3^\dagger$ \\ 
%\toprule
\hline
\hline
 \end{tabular}
\vspace*{-.2in}
\end{table}

\section{Possible baryon-number and/or lepton-number violating processes}
We now turn to the possible minimal scalar interactions that mediate
either baryon and/or lepton number violation but conserve SM gauge symmetries. 
The possible interactions, 
including as many as four distinct scalars, are enumerated in 
Table~\ref{table:Xprods}. The models labelled M1-M9 are those of Models 1-9, 
respectively, in Ref.~\cite{Arnold:2012sd}. A particular model contains terms that couple the
scalars to fermions and terms that couple the scalars to each other. We find
we must modify the scalar self-couplings of M2 and M7 in order to maintain
electric charge conservation for each term of the scalar self-interaction. Rather
than recapitulate M1-M9 
we simply summarize the detailed versions of the 
scalar forms enumerated in Table~\ref{table:Xprods}: 
\begin{alignat}{3}
&\lambda_1 X_5^{\alpha\alpha'}X_5^{\beta\beta'}X_7^{\gamma\gamma'}\epsilon_{\alpha\beta\gamma}\epsilon_{\alpha'\beta'\gamma'} \,, \quad \quad 
&&\lambda_2 [X_4^{\alpha \alpha' A} X_4^{\beta \beta' B}]_{0}
X_7^{\gamma\gamma'}\epsilon_{\alpha\beta\gamma}\epsilon_{\alpha'\beta'\gamma'} \,, \nonumber\\
&\lambda_3 X_7^{\alpha\alpha'}X_7^{\beta\beta'}X_8^{\gamma\gamma'}\epsilon_{\alpha\beta\gamma}\epsilon_{\alpha'\beta'\gamma'} \,, \quad\quad 
&& \lambda_4 X_{6\alpha}X_{6\beta}X_8^{\alpha\beta} \,, \nonumber \\ 
&\lambda_5 X_5^{\alpha\alpha'}X_5^{\beta\beta'}X_5^{\gamma\gamma'}X_2\epsilon_{\alpha\beta\gamma}\epsilon_{\alpha'\beta'\gamma'} \,, \quad \quad 
&&\lambda_6 X_4^{\alpha\alpha'A}X_4^{\beta\beta'B}X_4^{\gamma\gamma'C}X_2\epsilon^{ABC}\epsilon_{\alpha\beta\gamma}\epsilon_{\alpha'\beta'\gamma'} \,, \nonumber \\
&\lambda_7 [X_4^{\alpha\alpha'A}X_4^{\beta\beta'B}X_4^{\gamma\gamma'C}X^D_3 ]_{0}
\epsilon_{\alpha\beta\gamma}\epsilon_{\alpha'\beta'\gamma'} \,, \quad  
&&\lambda_8 X_7^{\alpha\alpha'} X_7^{\beta\beta'} X_7^{\gamma\gamma'} X_1^{\dagger} 
\epsilon_{\alpha\beta\gamma}\epsilon_{\alpha'\beta'\gamma'} \,, \nonumber \\
&\lambda_9 X_{6 \alpha} X_{6 \beta} X_{6 \gamma} X_1^{\dagger} \epsilon^{\alpha\beta\gamma} \,, 
\label{scalarself}
\end{alignat} 
where Hermitian conjugation is implied. 
The noted weak singlets follow from SU(2) Clebsch-Gordon 
coefficients~\cite{Patrignani:2016xqp}, so that 
\begin{equation}
[X_4^{\alpha \alpha' A} X_4^{\beta \beta' B}]_{0}
\equiv \frac{1}{\sqrt{3}}[X_4^{\alpha\alpha'1}X_4^{\beta\beta'3} 
+ X_4^{\alpha\alpha'3}X_4^{\beta\beta'1} - X_4^{\alpha\alpha'2}X_4^{\beta\beta'2}] \,  
\label{2sing}
\end{equation}
and 
\begin{eqnarray}
\!\!&& [X_4^{\alpha\alpha'A}X_4^{\beta\beta'B}X_4^{\gamma\gamma'C}X^D_3]_{0} 
\equiv \frac{1}{\sqrt{3}}\Bigg\{ 
\Big[
\sqrt{\frac{3}{5}}  \chi_4^{\alpha \alpha' 1} \chi_4^{\beta \beta' 1} \chi_4^{\gamma \gamma' 3} 
\nonumber \\
\!\!&&- (\sqrt{\frac{3}{20}} - \frac{1}{2}) 
\chi_4^{\alpha \alpha' 1} \chi_4^{\beta \beta' 2} \chi_4^{\gamma \gamma' 2} 
- 
(\sqrt{\frac{3}{20}} + \frac{1}{2})
\chi_4^{\alpha \alpha' 2} \chi_4^{\beta \beta' 1} \chi_4^{\gamma \gamma' 2} \nonumber \\
\!\!&&+ (\sqrt{\frac{1}{60}} - \frac{1}{2} + \sqrt{\frac{1}{3}}) 
\chi_4^{\alpha \alpha' 1} \chi_4^{\beta \beta' 3} \chi_4^{\gamma \gamma' 1} 
%\nonumber \\ \!\!&&
+ (\sqrt{\frac{1}{60}} + \frac{1}{2} + \sqrt{\frac{1}{3}}) 
\chi_4^{\alpha \alpha' 3} \chi_4^{\beta \beta' 1} \chi_4^{\gamma \gamma' 1} \nonumber \\
\!\!&& + (\sqrt{\frac{1}{15}} - \sqrt{\frac{1}{3}}) 
\chi_4^{\alpha \alpha' 2} \chi_4^{\beta \beta' 2} \chi_4^{\gamma \gamma' 1} 
\Big] \chi_3^3 
 + \Big[ ``1'' \leftrightarrow ``3''  \Big] \chi_3^1 \nonumber \\
% fix typo 11/28/18
\!\!&& - 
\Big[(\sqrt{\frac{3}{20}} + \frac{1}{2})(
\chi_4^{\alpha \alpha' 1} \chi_4^{\beta \beta' 2} \chi_4^{\gamma \gamma' 3} 
 + 
%(\sqrt{\frac{3}{20}} + \frac{1}{2}) 
\chi_4^{\alpha \alpha' 3} \chi_4^{\beta \beta' 2} \chi_4^{\gamma \gamma' 1}) 
\nonumber \\
\!\!&& + 
(\sqrt{\frac{3}{20}} - \frac{1}{2}) (
\chi_4^{\alpha \alpha' 2} \chi_4^{\beta \beta' 3} \chi_4^{\gamma \gamma' 1} 
 + 
%(\sqrt{\frac{3}{20}} - \frac{1}{2}) 
\chi_4^{\alpha \alpha' 2} \chi_4^{\beta \beta' 1} \chi_4^{\gamma \gamma' 3} ) 
\nonumber \\
\!\!&& -
(\sqrt{\frac{1}{15}} - \sqrt{\frac{1}{3}} )( 
\chi_4^{\alpha \alpha' 1} \chi_4^{\beta \beta' 3} \chi_4^{\gamma \gamma' 2} + 
% -
%(\sqrt{\frac{1}{15}} - \sqrt{\frac{1}{3}}) 
\chi_4^{\alpha \alpha' 3} \chi_4^{\beta \beta' 1} \chi_4^{\gamma \gamma' 2}) \nonumber \\
\!\!&& - 
(\sqrt{\frac{4}{15}} + \sqrt{\frac{1}{3}}) 
\chi_4^{\alpha \alpha' 2} \chi_4^{\beta \beta' 2} \chi_4^{\gamma \gamma' 2} 
\Big] \chi_3^2 \Bigg\} \,,
\end{eqnarray}
where ``$\,\text{`1'} \leftarrow \text{`3'}\,$'' 
denotes the expression found by exchanging 1 and 3 superscripts. 
%from that in the previous square brackets. 
%%\end{onecolumn} 
Turning to the $\DeltaLm=2$  
models in Table~\ref{table:Xprods}, 
we find
\begin{alignat}{5}
&\lambda_A X_8^{\alpha\alpha'}(X_7^{\alpha\alpha'})^\dagger X_1 \,,\quad  
&&\lambda_B [X_3^A X_4^{\alpha \alpha' B}]_{0} (X_7^{\alpha \alpha'})^\dagger \,,\quad  
&&\lambda_C [X_3^A (X_4^{\alpha \alpha' B})^\dagger ]_{0} X_8^{\alpha \alpha'} \,, \nonumber \\ 
&\lambda_D X_5^{\alpha\alpha'}(X_7^{\alpha\alpha'})^\dagger X_2 \,,\quad  
&&\lambda_E X_8^{\alpha\alpha'}(X_5^{\alpha\alpha'})^\dagger X_2 \,,\quad  
&&\lambda_F X_2 X_2 X_1^\dagger  \,, \nonumber \\ 
&\lambda_G [X_3^A X_3^B]_0 X_1^\dagger \,,
\end{alignat}
whereas for the remaining baryon-number-violating models, we have 
\begin{alignat}{3}
&\lambda_{10} X_7^{\alpha\alpha'}X_8^{\beta\beta'}X_8^{\gamma\gamma'}X_1
\epsilon_{\alpha\beta\gamma}\epsilon_{\alpha'\beta'\gamma'} \,,\quad
&&\lambda_{11} X_5^{\alpha\alpha'}X_5^{\beta\beta'}
[X_4^{\gamma \gamma' A} X_3^{B}]_{0} 
\epsilon_{\alpha\beta\gamma}\epsilon_{\alpha'\beta'\gamma'} \,, 
\nonumber \\
&\lambda_{12} X_5^{\alpha\alpha'} X_5^{\beta\beta'} X_8^{\gamma\gamma'} X_1
\epsilon_{\alpha\beta\gamma}\epsilon_{\alpha'\beta'\gamma'} \,, \quad
&&\lambda_{13} [X_4^{\alpha\alpha'A} X_4^{\beta\beta'B}]_0 X_5^{\gamma\gamma'}X_2
\epsilon_{\alpha\beta\gamma}\epsilon_{\alpha'\beta'\gamma'} \,, \nonumber \\ 
&\lambda_{14} X_4^{\alpha\alpha'A} X_4^{\beta\beta'B} X_3^C X_5^{\gamma\gamma'}
\epsilon^{ABC}
\epsilon_{\alpha\beta\gamma}\epsilon_{\alpha'\beta'\gamma'} \,, \quad 
&&\lambda_{15} [X_4^{\alpha\alpha'A} X_4^{\beta \beta' B}]_{0} X_8^{\gamma \gamma'} X_1
\epsilon_{\alpha\beta\gamma}\epsilon_{\alpha'\beta'\gamma'} \,, \nonumber \\ 
&\lambda_{16} [X_4^{\alpha\alpha'A} X_3^B ]_{0} X_7^{\beta \beta'} X_8^{\gamma \gamma'}
\epsilon_{\alpha\beta\gamma}\epsilon_{\alpha'\beta'\gamma'} \,, \quad
&&\lambda_{17} X_5^{\alpha\alpha'} X_7^{\beta \beta'} X_7^{\gamma \gamma'} X_2^\dagger 
\epsilon_{\alpha\beta\gamma}\epsilon_{\alpha'\beta'\gamma'} \,, \nonumber \\ 
&\lambda_{18} [X_4^{\alpha \alpha' A} (X_3^{B})^\dagger ]_0 
X_7^{\beta \beta'} X_7^{\gamma \gamma'} 
\epsilon_{\alpha\beta\gamma}\epsilon_{\alpha'\beta'\gamma'} \,, 
\end{alignat}
and Hermitian conjugation is implied throughout. Models with $X_2$ and
$X_6$ couple to leptons and quarks of different generations. 
Only models M1, M2, and M3 can produce $\nnbar$ oscillations, 
though these models do not generate all the low-energy 
effective operators expected 
if SM gauge symmetry holds~\cite{Rao:1982gt,Caswell:1982qs,Gardner:2017szu}. 
In particular, 
we find that M1 yields the operator $({\cal O}_{2})_{RRR}$, M2 yields $({\cal O}_{1})_{LLR}$ and
$({\cal O}_{2})_{LLR}$~\cite{Rao:1982gt}, though an operator relation combines these to 
$({\cal O}_{3})_{LLR}$~\cite{Caswell:1982qs} 
and M3 yields $({\cal O}_{1})_{RRR}$. 
An operator of form $({\cal O}_3)_{LLR}$ 
can also appear~\cite{Rao:1982gt,Caswell:1982qs}, 
but it is not generated 
in the minimal scalar-fermion models we consider. 

Only models A, B, and C can produce $\pi^-\pi^- \to e^-e^-$ decay, though B and C
can also yield a weak isospin triplet of $\DeltaLm=2$ processes. These models all correspond
to the second case of decay topology ``T-II-3'' in Ref.~\cite{Bonnet:2012kh}, as that
decomposition considers the scalars' electric and color charge only. 
At energies below the $X_i$ mass scale, 
model A generates the operator combination ${\cal O}_{3+}^{++} - {\cal O}_{3-}^{++}$, 
whereas models B and C generate linear combinations 
of ${\cal O}_{2\pm}^{++}$~\cite{Prezeau:2003xn}. 

\section{Phenomenology}
The models of Table~\ref{table:Xprods} possess a rich array of possible 
$\DeltaBm =2$ and $\DeltaLm =2$ processes. They also reveal the 
possibility of scattering-mediated $\DeltaBm=2$ processes, 
which we term ``conversion'' modes~\cite{Gardner:2016wov, Gardner:2017szu}, 
and we show some of the more experimentally 
accessible ones in Table~\ref{table:pheno}. As they are 
%. Those 
mediated by mass dimension 12 operators, they do not break B-L~\cite{Kobach:2016ami}. 
%baryon-number violation through scattering, that is, of nucleon-antinucleon
%conversion
Other models show additional features. Models D and E support 
$\pi^- \pi^0 \to e^-  \nu_{\mu,\tau}$ and $\pi^- \pi^0 \to \mu^- \nu_{e}$, 
whereas F supports $\mu^- \to e^- e^+ e^- \bar\nu_e \bar\nu_\mu$ and G supports 
$e^+ e^- \to e^+ e^- \bar\nu_e \bar\nu_e$. Models M8 and M18 can mediate 
$nn \to \pi^+\pi^+ e^- e^-$ decay, and finally M17 and M18 can yield 
$e^+ n \to {\bar \Delta}^+ \nu_{\mu,\tau}$ and $e^+ n \to {\bar \Delta}^+ \nu_{e}$
processes, respectively. We review the existing experimental constraints
on the scalars we have considered in Sec.~\ref{obs}.

\begin{table}[tb]
 \caption{Suite of $\DeltaBm=2$ and $\DeltaLm=2$ processes generated by the models
of Table~\ref{table:Xprods}, focusing on states with first-generation matter. 
The $(*)$ superscript indicates that a weak isospin triplet of $\DeltaLm =2$ processes can
appear, namely $\pi^0\pi^0 \to \nu\nu$ and $\pi^-\pi^0 \to e^- \nu$.
Models M7, M11, M14, and M16 also support $\nu n \to \bar{n} {\bar \nu}$, revealing that 
cosmic ray neutrinos could potentially mediate a $\DeltaBm=2$ effect. }
% \vspace*{-.1in}
 \label{table:pheno}
\begin{tabular}[t]{|c|c|c|c|c|}
% \toprule 
\hline
\hline
$n\bar{n}$ & $\pi^-\pi^-\rightarrow e^-e^-$ & $e^-p\rightarrow \bar{\nu}_{\mu,\tau}\bar{n}$ & 
$e^-p\rightarrow \bar{\nu}_{e}\bar{n}/ e^+ {\bar p}$ 
& $e^-p\rightarrow e^+\bar{p}$ \\ \hline 
\hline
M1 & A & M5 & M7 & M10 \\
M2 & B$^{(*)}$ & M6 & M11 & M12 \\
M3 & C$^{(*)}$ & M13 & M14 & M15 \\
 & & & M16 &  \\
%\toprule
\hline
\hline
 \end{tabular}
\vspace*{-.2in}
\end{table}

\section{Connecting $\DeltaBm =2$ to $\DeltaLm =2$ processes with new physics}
\label{BNVLNV}
The scalar-fermion models that yield $\nnbar$ oscillations can differ
in just one scalar from models that generate $\DeltaLm=2$ processes and indeed
$0\nu \beta\beta$ decay. We now discuss how an observed pattern of 
baryon-number-violating conversion modes, all accessible through e-d scattering, 
can determine both the $\nnbar$ model and whether such an additional scalar exists. 
To distinguish the possibilities, detecting both the appearance of an 
antinucleon and the electric charge of a final-state charged lepton is necessary. 
For context, we note that 
M3 has scalar content $X_7 X_7 X_8$ but A has $X_1 X_8 X_7^\dagger$, that 
M2 has $X_4 X_4 X_7$ but B has $X_3 X_4 X_7^\dagger$, that M1 has $X_5 X_5 X_7$ but
D has $X_5 X_7^\dagger X_2$ --- and finally that C has $X_3 X_8 X_4^\dagger$, where 
the Hermitian conjugate is implied here and henceforth. 
If $\nnbar$ oscillation occurs, then $e^- n \to e^- {\bar n}$ can appear also, 
if the mediating operator is not $({\cal O}_1)_{RRR}$~\cite{Gardner:2017szu}. Thus 
the latter process acts as a diagnostic of the possible $\nnbar$ model. 

Possible patterns of $\DeltaBm=2$ discovery are shown for the different $\nnbar$
models in Table~\ref{table:test}. Model M3 can connect to model A through 
models M8, containing $X_7 X_7 X_7 X_1^\dagger$, and M10, containing 
$X_7 X_8 X_8 X_1^\dagger$, though only the latter can be probed through $e-p$ scattering,
as shown in Table~\ref{table:pheno}. Consequently, observing a $n-{\bar n}$ oscillation and 
the process $e^- p \to e^+ {\bar p}$  
in the absence of $e^- n \to e^- n$ and 
$e^- p \to {\bar \nu}_X {\bar n}$ 
would point to model M3 and the existence of $X_1$. With these observations, we then would have
experimental evidence for all the new degrees of freedom in model A. 
Thus model A, with its minimal scalar interaction, 
should also exist because there would be no reason that it should not. 
This thinking was promoted by Gell-Mann in the early days of the quark model: 
%Gell-Mann's principle that 
that what is not forbidden is compulsory~\cite{Gell-Mann:1956iqa}. 
We can also make a connection to model A by drawing a Feynman diagram for 
$\pi^-\pi^- \to e^- e^-$   
utilizing the interactions of models M3 and M10; this is illustrated in 
Fig.~1. However, 
this suggests that the rate for $\pi^-\pi^- \to e^- e^-$, although nonzero, 
would also be vanishingly small. 
We do not think this latter conclusion is necessary
because model A itself is minimal. 

Other connections are possible and can be distinguished by the pattern of 
observables shown in Table~\ref{table:test}. 
Observing a $\nnbar$ oscillation 
and $e^- n \to e^- n$ would reveal that either M2 or M1 operate, though 
the pattern of $\DeltaBm=2$ $e-p$ processes shown can also discriminate
between the three $n$-${\bar n}$ models. 
Model M2 is associated with the interaction $X_4 X_4 X_7$,
and model B is 
%, that yields $\pi\pi\to e^- e^-$, is
associated with $X_3 X_4 X_7^\dagger$.
Minimal models with a four-scalar interaction that connect them are
M7, with $X_4 X_4 X_4 X_3$, 
or M18, with  $X_4 X_7 X_7 X_3^\dagger$, though only M7 
can generate a process with an $e^- p$ or $e^- n$ initial state.
Model M7 can give rise to $e^- p \to e^+ {\bar p}$ and 
$e^- p \to {\bar \nu}_e {\bar n}$.
Note that a Feynman diagram utilizing M2 and M7 can generate 
model B and $\pi^-\pi^- \to e^- e^-$. 
In contrast, 
model M1 is associated with $X_5 X_5 X_7$, and model D, that
yields lepton number and flavor violation, is associated with $X_5 X_2 X_7^\dagger$.
Here the minimal four-scalar models are M5, with $X_5 X_5  X_5 X_2$, or M17,
with $X_5 X_7 X_7 X_2^\dagger$, though only M5 can give rise to 
$e^- p \to {\bar \nu}_{\mu,\tau} {\bar n}$.  
Here a Feynman diagram utilizing M1 and M5 generates model D and, e.g.,  
$\pi^-\pi^0 \to e^- \nu_{\mu,\tau}$. The two sets of possibilities can be
distinguished as follows. 
If $e^- p \to {\bar \nu}_X {\bar n}$ and $e^- p \to e^+ {\bar p}$ are both observed, 
in addition to a $\nnbar$ oscillation, 
then this would point to the existence of $X_3$ and thus models M2 and B. 
However, 
if $e^- p \to e^+ {\bar p}$ is instead absent, this would point to the existence of
$X_2$ and thus models M3 and D. 
Note that 
the various model possibilities cannot combine to show that {\em only} 
$X_8$ exists, even if the noted $\DeltaBm=2$ processes are observed, so that we cannot 
show that model C operates. 
The observed patterns would establish 
the existence of $\DeltaLm=2$ processes from new short-distance physics, but the 
connections we argue would not exclude the latter possibility if no $\DeltaBm=2$ 
processes were observed. 
\begin{figure}[h]
%\vspace{-2.5cm}
%\begin{center}
\centering
%\hspace{-1.75cm}
%\includegraphics[scale=0.50,angle=-90]{Feynman_comb_03.pdf}
%\includegraphics[scale=0.50,angle=-90]{Feyn06_03.pdf}
\includegraphics[scale=0.70]{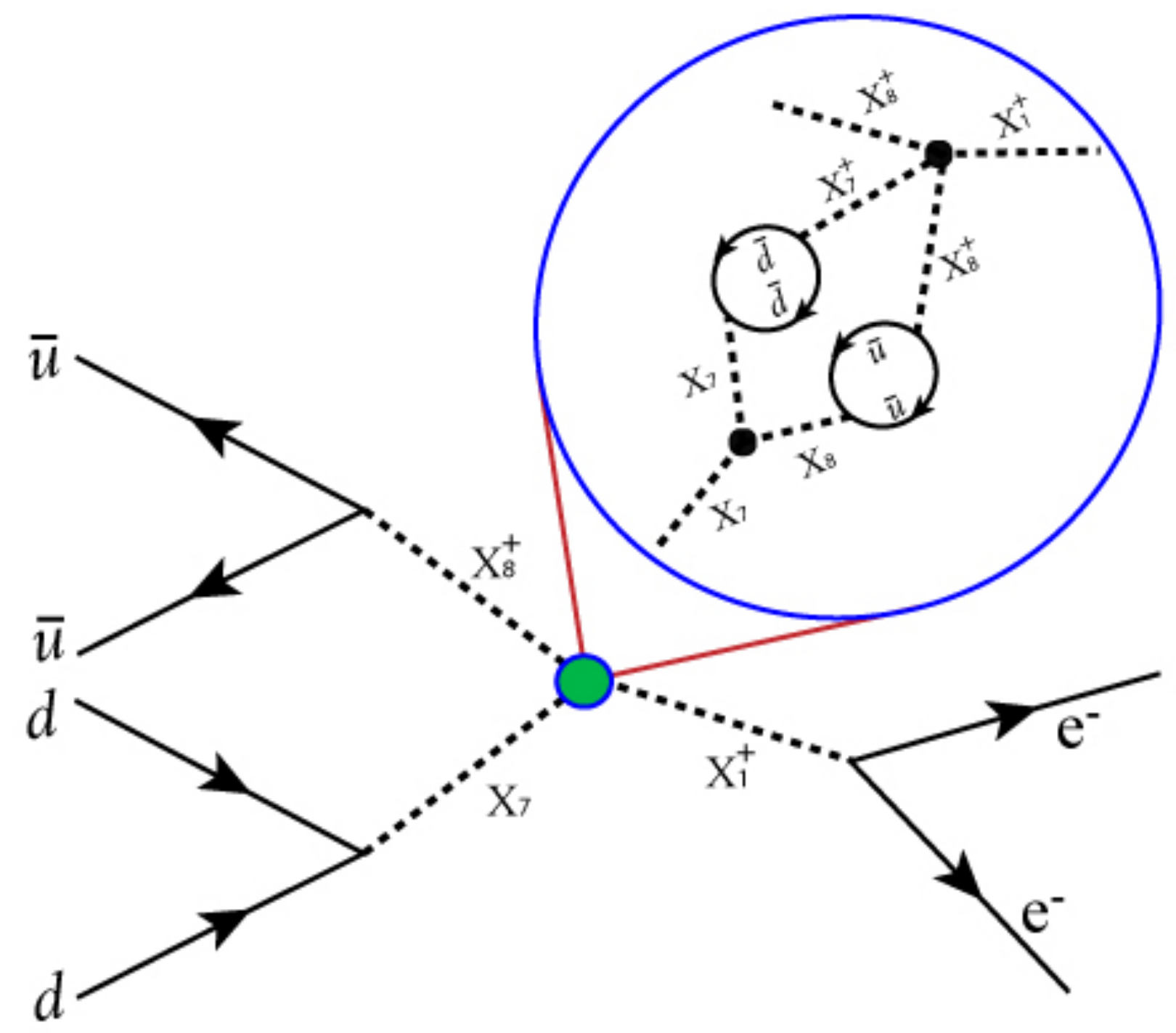}
%\includegraphics[width=\textwidth, clip = true, trim = 2mm 16mm 0 0]{nnbarconv_link_zeronu_fig.pdf}
%\vspace{1.5cm}
\caption{A Feynman diagram for $\pi^-\pi^- \to e^- e^-$ in model A
utilizing the interactions of models M3 and M10.}
%\end{center}
%\vspace{2.5cm}
\end{figure}
%\vspace{2.5cm}
%, recalling Gell-Mann's 
%totalitarian principle, 
%

The connections we consider exist regardless of whether the neutrino also
has a Dirac mass. Note that if $\nu_R$ fields existed in the low-energy theory, 
not only could
the neutrino have a Dirac mass, but the $X_6$ scalar could also induce proton
decay. Thus this possibility would rule out models M4, M9, M19-M21, but they are not
pertinent to our arguments. 
We also note that independent constraints on $X_7$ and $X_8$ 
can be had from studies 
of $K\bar K$ and $D\bar D$ mixing, respectively. 
Thus the discovery of new physics in $D\bar D$ mixing could also help
anchor evidence for Model C and $0\nu \beta\beta$ decay from new short-distance physics. 

\begin{table}[tb]
 \caption{
Possible patterns of $\DeltaBm=2$ discovery and their interpretation 
in minimal scalar-fermion models. Note that only $\nnbar$ oscillations 
and $e^-  n \to e^-  {\bar n}$ break B-L symmetry and that the pertinent
conversion processes can be probed through electron-deuteron scattering. 
The latter are distinguished by the electric charge of the final-state lepton
accompanying nucleon-antinucleon annihilation. 
Note that the $0\nu \beta\beta$ query refers specifically to 
the existence of $\pi^- \pi^- \to e^- e^-$ 
from 
new, short-distance physics. 
Note that we can possibly establish model D and $\DeltaLm=2$ violation, 
but that model does not give rise to $\pi^- \pi^- \to e^- e^-$. In contrast
we cannot establish $X_8$ alone and thus cannot establish model C. 
} 
% \vspace*{-.1in}
 \label{table:test}
\begin{tabular}[t]{cccccc}
\hline
\hline
Model & $n{\bar n}$? & $e^- n \to e^- {\bar n}$?  & $e^- p \to {\bar \nu_X} {\bar n}$? 
& $e^- p \to e^+ {\bar p}$? & $0\nu \beta\beta$ ? \\
\hline
\hline
M3 & Y & N & N & Y &  Y [A]  \\
M2 & Y & Y & Y & Y &  Y [B]  \\
M1 & Y & Y & Y & N &  ? [D]  \\
--  & N & N & Y & Y &  ? [C?] \\
\hline
\hline
 \end{tabular}
\vspace*{-.2in}
\end{table}

\section{Observability}
\label{obs}
The non-observation of $\nnbar$ oscillations~\cite{BaldoCeolin:1994jz,Abe:2011ky} 
can be interpreted as a limit on the neutron's Majorana mass of 
$2\times 10^{-33}$ GeV at 90\% CL~\cite{Abe:2011ky}, with 
greatly improved sensitivity anticipated 
at a new 
experiment proposed for the 
European Spallation Source~\cite{Milstead:2015toa}.
Such limits do not preclude the observation of processes associated with the
dimension-12 operators we have considered, because
different scalars can have different masses.  
The scalar self-interactions we consider do not select a particular mass scale; 
rather, the allowed masses and couplings should be determined from experiment, 
as in hidden-sector searches~\cite{Alexander:2016aln}. We find that 
the various $e-p$ processes we have considered should be appreciable if the scalars
possess masses of ${\cal O}(1-10\,{\rm GeV})$. 
Existing collider 
constraints on color-sextet scalars (of ${\cal O}(500\,{\text{GeV}})$ 
with ${\cal O}(1)$ couplings) 
come from studies of $t$-quark final 
states~\cite{Mohapatra:2007af,Cacciapaglia:2015eqa,Aad:2015gdg,Aad:2015kqa}, and
flavor-physics constraints, while more severe, also involve
second- and third-generation quark-scalar 
couplings~\cite{Babu:2006xc,Babu:2006wz,Babu:2008rq,Arnold:2012sd,Babu:2013yca,Fortes:2013dba}. 
Thus these constraints are not really pertinent to our case. However, 
there are also limits specific to scalars that couple to first-generation fermions; 
here we summarize findings that we plan to report on detail elsewhere~\cite{svgxy19}. 
Severe limits on $pp \to e^+ e^+$ in $^{16}$O have recently been reported by 
the Super-Kamiokande collaboration~\cite{Sussman:2018ylo}. 
Such limits must be interpreted
carefully because conventional physics can act to  make the spontaneous process impossible,
regardless of whether new physics is present. It has been claimed that 
earlier studies already limit the scalar mass scale to no less 
than $1.6$ TeV~\cite{Bramante:2014uda}, though 
that analysis neglects the role of Coulomb repulsion in the $p p$ initial state. 
Its inclusion should weaken that bound by orders of magnitude. 
In addition, $e^- p \to e^+ {\bar p}$ from K-shell capture in $^{16}$O would not occur
spontaneously because only the initial lepton can be in an atomic bound state. 
There are also astrophysical limits
on hydrogen-antihydrogen ($H-{\bar H}$) oscillation 
from attributing a measured excess of gamma radiation to the annihilation of 
%measured by the Fermi LAT to the annihilation
${\bar H}$ atoms from $H-{\bar H}$ 
oscillations~\cite{Grossman:2018rdg}. That analysis neglects 
Galactic magnetic fields, which act to make the energy of  $H$ and ${\bar H}$ unequal, 
quenching the oscillation probability. Magnetic fields of about 1 nT have been 
established in cold, HI clouds~\cite{crutcher}, and magnetic fields of no less
than 0.1 nT exist in the warm interstellar medium~\cite{haverkorn}. 
Thus we believe that cold, HI regions continue 
to drive the assessed $H-{\bar H}$ oscillation limit as estimated in 
Ref.~\cite{Grossman:2018rdg}. 
Computing the $H-{\bar H}$ energy splitting, we estimate the oscillation limit
to be weakened by a factor of $10^8$. 
Collider searches for events with same-sign 
dileptons and multiple jets at the center-mass energies of 
$\sqrt{s}=$7, 8,  and 13 TeV have been performed 
by the CMS collaboration~\cite{Chatrchyan:2012sa,Chatrchyan:2013fea,Khachatryan:2016kod}.  
Due to backgrounds 
from $b$-hadron decays, they reject same-sign dilepton events with an invariant mass of 
less than 8 GeV~\cite{Chatrchyan:2012sa}. 
Thus, these collider constraints do not exclude possibility of models with scalars that
couple to dileptons with masses that are less than 8 GeV. 
%energy scale below 
%Since only first-generation leptons are involved, constraints 
%set from processes, such as heavy lepton decays, muonium-antinuonium oscillations, 
%and ($\mu\rightarrow e\gamma$)-type process, on scalars that couples to dileptons 
%are automatically evaded. 
%Finally, we find that constraints set from electron anomalous magnetic momenta 
%can be easily satisfied as long as 
%the new scalar is much heavier than electrons. 
With these various refinements in place we believe that scalars 
with masses of ${\cal O}(1-10\,{\rm GeV})$ 
are a viable possibility. 

%Although there is no such invariant mass cut for dijets, 
%since the transverse momenta cuts for jets are always at least 2 times bigger than ones for leptons, 
%it is reasonable to expect that there exists an implicit invariant mass cuts for dijets, too,  
%and it should be bigger than 8 GeV. Thus, these collider constraints do not exclude possibility of new physics with 
%energy scale below 8 GeV. Since only first-generation leptons are involved, constraints 
%set from processes, such as heavy lepton decays, muonium-antinuonium oscillations, 
%and ($\mu\rightarrow e\gamma$)-type process, on scalars that couples to dileptons are automatically evaded. 
%Finally, we find that constraints set from electron anomalous magnetic momenta can be easily satisfied as long as 
%the new scalar is much heavier than electrons. 
%With these various refinements in place we believe that scalars 
%with masses of ${\cal O}(1-8\,{\rm GeV})$ 
%to be a viable possibility. 

Models that support 
$e^- p \to e^+ {\bar p}$ 
have low-energy operators whose quark parts 
correspond to those found in $\nnbar$ oscillations under $u \leftrightarrow d$ exchange. 
Exploiting this and a MIT bag model~\cite{Chodos:1974je,Chodos:1974pn}
computation of 
$\langle {\bar n} | ({\cal O}_{1,2})_{LLL} | n \rangle$~\cite{Rao:1982gt,Yan:2017eky}
yields 
%1.5 \times 10^{-5}
\begin{equation}
\!\!\! \sigma \sim 
1.5 \times 10^{-4}
|g_4^{11}|^6 |\lambda_7|^2 |g_3^{11}|^2 
\bigg(\frac{5\, \text{GeV}}{M_{X_4}}\bigg)^{\!12} \!
\bigg(\frac{1\, \text{GeV}}{M_{X_3}}\bigg)^{\!4} \!
\text{ab}
\label{cross}
\end{equation}
in model M7 for an electron beam energy of 155 MeV with a fixed target~\cite{Becker:2018ggl}. 
Model M7 contains scalars distinct from those that generate
$\nnbar$ oscillations, and existing phenomenological analyses allow scalars in the
${\cal O}(1-10\,{\rm GeV})$ mass range to appear. 
%the size of the $e^- p \to e^+ {\bar p}$ cross section 
%is not limited by existing empirical limits on $|\Delta B|=2$ processes. 
%A broad range of possible scalar masses and couplings exists, and 
The experimental searches 
we propose, given Eq.~(\ref{cross}) and the established accelerator 
and target capacities we have collected in 
Ref.~\cite{Gardner:2017szu}, can discover or constrain them. 

\section{Summary} 
We have considered different physical processes that could reveal $\DeltaBm=2$ violation, 
both $\nnbar$ oscillation and conversion, and we 
have considered their interrelationships within 
minimal scalar-fermion models that support $\DeltaBm =2$ processes without proton decay. 
To realize this we have extended the models of Ref.~\cite{Arnold:2012sd} to include
all possible minimal-scalar models that satisfy SM gauge invariance. 
Three distinct scalars are required to realize neutrinoless double $\beta$ decay
in these models, and Ref.~\cite{Arnold:2012sd} considered no more than two distinct
scalars. Moreover, we have shown how the patterns of observation of particular
$\DeltaBm=2$ processes would speak to the existence of particular new scalars within these
models, and we have employed Gell-Mann's totalitarian principle~\cite{Gell-Mann:1956iqa} 
%totalitarian principle 
%notion that what is not forbidden is compulsory 
 to invoke the new 
combination of these scalars needed to predict the existence of $\pi^-\pi^- \to e^- e^-$ and thus
of neutrinoless double $\beta$ decay, though the latter connection also follows
from a Feynman diagram approach once the particular $\DeltaBm =2$ processes 
are observed. Thus, finally, we conclude that 
the observation of particular $\DeltaBm =2$ processes 
could be used to infer the existence of a $\DeltaLm =2$ process,
$0\nu \beta\beta$ decay in nuclei, speaking 
to the Majorana nature of the neutrino and to 
new dynamics at accessible energy scales.

\section*{Acknowledgments}
We acknowledge partial support from the Department
of Energy Office of Nuclear Physics under 
contract DE-FG02-96ER40989. 

\section*{References}

\bibliography{majornnbar_plb}

\end{document}